\documentclass[a4paper,11pt]{article}
\usepackage{pos}
\usepackage{wrapfig}
\usepackage{subfigure}
\usepackage{float}
\usepackage[compact]{titlesec} 

\title{Performance of the muon track reconstruction with the Baikal-GVD neutrino telescope}
 \ShortTitle{Baikal-GVD muon track reconstruction}

\author*[a]{Grigory Safronov}

\affiliation[a]{Institute for Nuclear Research of the Russian Academy of Sciences,\\
  Prospekt 60-letiya Oktyabrya 7a, Moscow, Russia}

\forColl{Baikal-GVD} 

\emailAdd{grigorybs@gmail.com}

\abstract{Baikal-GVD is a km$^3$-scale neutrino telescope being constructed in Lake Baikal. Muon and partially tau (anti)neutrino interactions near the detector through the W$^{\pm}$-boson exchange are accompanied by muon tracks. Reconstructed direction of the track is arguably the most precise probe of the neutrino direction attainable in Cerenkov neutrino telescopes. Muon reconstruction techniques adopted by Baikal-GVD are discussed in the present report. Performance of the muon reconstruction is studied using realistic Monte Carlo simulation of the detector. The algorithms are applied to real data from Baikal-GVD and the results are compared with simulations. The performance of the neutrino selection based on a boosted decision tree classifier is discussed.}

\FullConference{37$^{\rm{th}}$ International Cosmic Ray Conference (ICRC 2021)\\
		July 12th -- 23rd, 2021\\
		Online -- Berlin, Germany}

\begin{document}

\maketitle

\section{Introduction}
\begin{wrapfigure}[17]{r}{0.6\textwidth}
\centering
  \includegraphics[width=0.55\textwidth]{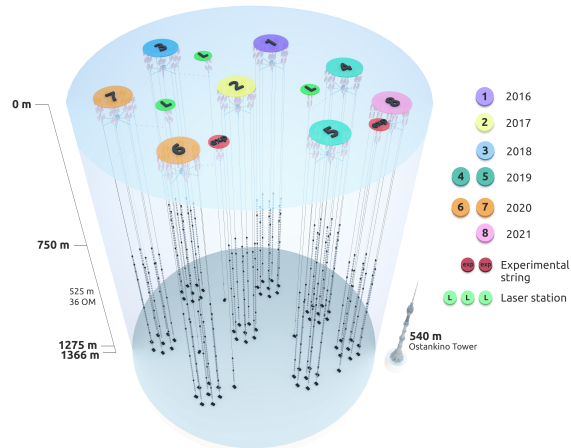}
  \caption{General view of Baikal-GVD detector in season 2021. Yearly progression of detector deployment is shown in legend. Technological strings carrying lasers and experimental strings with the new DAQ system are also shown.}
  \label{fig:det}
\end{wrapfigure}

The Baikal-GVD experiment site is located in the southern part of Lake Baikal $\sim$4 km away from shore where the lake depth is nearly constant at 1366-1367 m. The telescope consists of independent structural units - clusters. After the winter expedition in February-April 2021 the detector includes 8 clusters with the sum effective volume of $\sim$0.4 km$^{3}$ (Fig.~\ref{fig:det}). Muon and partially tau (anti)neutrino interactions in the vicinity of the detector through the W-boson exchange produce the muon track extending to large distances due to muon propagation range. Reconstruction techniques applied to events with sufficient track length within the sensitive detector volume allow to measure the direction of the neutrino with a sub-degree precision, thus making the muon channel the best for the point source association. We have developed a fast algorithm optimised for low-energy muon reconstruction based on $\chi^{2}$ minimisation and applied it to first few months of the season 2019, selecting a sample of 44 tracks dominated by atmospheric neutrino -induced muons \cite{nuatm_arxiv}. The fast reconstruction algorithm along with the neutrino selection method were deployed into the automatic data processing framework in 2021 identifying low-energy neutrino candidate events in Baikal-GVD data with the delay of few hours \cite{bairsh_auto}. In present report we discuss the extension of the algorithm towards high-energy muon reconstruction. In this extension along with muon detection efficiency improvement in wide energy range we introduce the energy measurement algorithm. Angular and energy resolution of the reconstruction are assessed using the realistic Monte Carlo (MC) datasets. Reconstruction is applied to 2019 data and data-MC comparisons are performed. An efficient neutrino selection algorithm based on boosted decision tree (BDT) classifier is presented. The paper is organised as follows: in chapter 2 the detector and data taking conditions are described along with datasets used, in chapter 3 the muon direction and energy reconstruction algorithms and their performance are discussed, in chapter 4 an efficient neutrino selection algorithm is discussed, chapter 5 contains the summary of presented results. 

\section{Baikal-GVD detector and used datasets}
Each cluster of the telescope consists of 8 strings each carrying 36 optical modules (OM) and calibration systems. OMs are located at depths from 750 to 1275 m with a vertical step of 15 m. Cluster radius is 60 m and  horizontal distance between central strings of neighbouring clusters is $\sim$300 m.
Baikal-GVD optical module is a glass sphere which hosts photomultiplier tube (PMT) Hamamatsu R7081-100 with a 10-inch hemispherical photocathode oriented towards the lake floor along with various sensors, readout and high-voltage control electronics, and calibration LED. OM readout is organised in sections each including 12 OMs. Readout controllers are housed in a section's Central Module (CM) in which digitisation of PMT signal occurs with a step of 5 ns. Time of signal arrival and charge deposited in PMT is derived from the pulse shape analysis. CM generates local trigger signal if pulses with deposited charge above adjustable thresholds $Q_{high}$ and $Q_{low}$ coincident within 100 ns time window are found in adjacent modules within one section. Typical values for $Q_{high}$ are 3-5 photoelectrons (p.e.) and 1-2 p.e. for $Q_{low}$ depending on the cluster and season. If a local trigger is generated within one of CM an event frame of 5~$\mu$s is read out from all CMs of the cluster and sent to the shore center via optoelectric cable attached to a central cluster control module. Data are further transmitted to JINR (Dubna) for event reconstruction and long-term storage. Data from different clusters are collected independently and merged into multi-cluster events in Dubna. Separate single-cluster and multi-cluster event datasets are available for offline analysis.

Each string carries a system of acoustic modems for the precise OM positioning. OM coordinates are reconstructed with the precision better than 20~cm \cite{acoustic}. Time calibration is performed using OM integrated LEDs and LED matrix modules which produce light flashes propagating to up to 100 m radius. A system of three technological strings located between clusters (Fig.~\ref{fig:det}) carries 5 dedicated lasers producing isotropical flashes which are used for intercluster calibration and water properties monitoring. The time calibration precision presently achieved is $\sim$2.5 ns~\cite{timecal}. The water light absorption length at Baikal-GVD site amounts to $\sim$22~m. The feature of the detector location is seasonally varied water chemiluminescence which dominates the PMT noise rate. Noise pulses at the level of 1 p.e. are produced at the rate of 20-50 kHz in quiet period in April - June and up to > 100 kHz at topmost OMs for some periods in the rest of the year \cite{noise}. 

Present report uses single-cluster data and MC samples while the multi-cluster data analysis is waiting for the final decision on data calibrations. At the same time the fast reconstruction performance was validated on multi-cluster MC~\cite{atm_neutrino}. For the reconstruction optimisation and performance studies we have used data and MC corresponding to quiet period of season 2019. Runs between April 1st and June 30th 2019 longer than an hour and passing good data quality criteria were selected. Total single cluster livetime of the data sample is 326 days with 63-72 days per cluster. Dedicated MC samples with realistic noise rate and charge deposition simulation using channel-wise trigger threshold values and taking into account dead channels were used. Signal samples are represented by atmospheric neutrino sample incorporating Bartol flux model~\cite{bartol} and by isotropic neutrino sample with $\sim E^{-2}$ spectrum and energy range $100$~GeV~$< E_\nu < 1$~ PeV referred to as high-energy neutrino sample. A simple neutrino generator code based on conventional cross-sections is used to simulate (anti)neutrino-nuclei interactions with water while muons are propagated through the instrumented volume using MUM v1.3u \cite{mum}. The background sample represented by atmospheric muon bundle sample is generated using CORSIKA 5.7 \cite{cors}. The energy of the cosmic ray primary particle was varied from 240 GeV to 2 PeV, generated events were reused and passed through the randomly rotated detector to enhance the statistics. In total $\sim$643 days of single custer livetime of atmospheric bundle MC sample were available for the studies.

\section{Muon reconstruction algorithms and their performance}

The muon track reconstruction is performed in two stages. At first stage the collection of PMT hits produced by Cerenkov light from the muon track is identified this is referred to as hit finder algorithm. At the second stage the track is reconstructed using minimisation algorithm, this is referred to as track fitter algorithm. In fast reconstruction~\cite{nuatm_arxiv} hit finder algorithm PMT pulses are clustered using the time and distance constraints with respect to the seed pulse with high charge deposition. A preliminary muon track direction estimation is performed as follows: 
\begin{equation}
\vec{R} = \sum_{ij \, (t_j>t_i)} w_{ij}(\vec{R_{j}}-\vec{R_{i}}), \;\;\;\; w_{ij}=q_{i}+q_{j}, 
\end{equation}
where $i$ and $j$ are pulses ordered in time and located at different strings and $q_i$ and $q_j$ are respective PMT charge depositions. Time and coordinates of the track are defined using pulse time and coordinates for the OM with largest charge deposition. The collection of hits is cleaned using time, distance and probability constraints with respect to the preliminary muon track iteratively in gradually tightening set of cuts. As a result of this procedure the collection of pulses with average noise contribution at the level of 1\% is selected. At the track fitter stage the muon track parameters are reconstructed by means of minimisation of the quality function $Q=\chi^{2}(t) + w*f(q,r)$ where $\chi^{2}(t)$ is the chi-square sum of time residuals with respect to direct Cerenkov light from the muon, $f(q,r)$ is the sum of products of charges deposited in OMs and their distances from the track and $w$ is the relative weight of the second term. 

\begin{figure}[ht]
\vspace{-10pt}
\centering
  \includegraphics[width=0.9\textwidth]{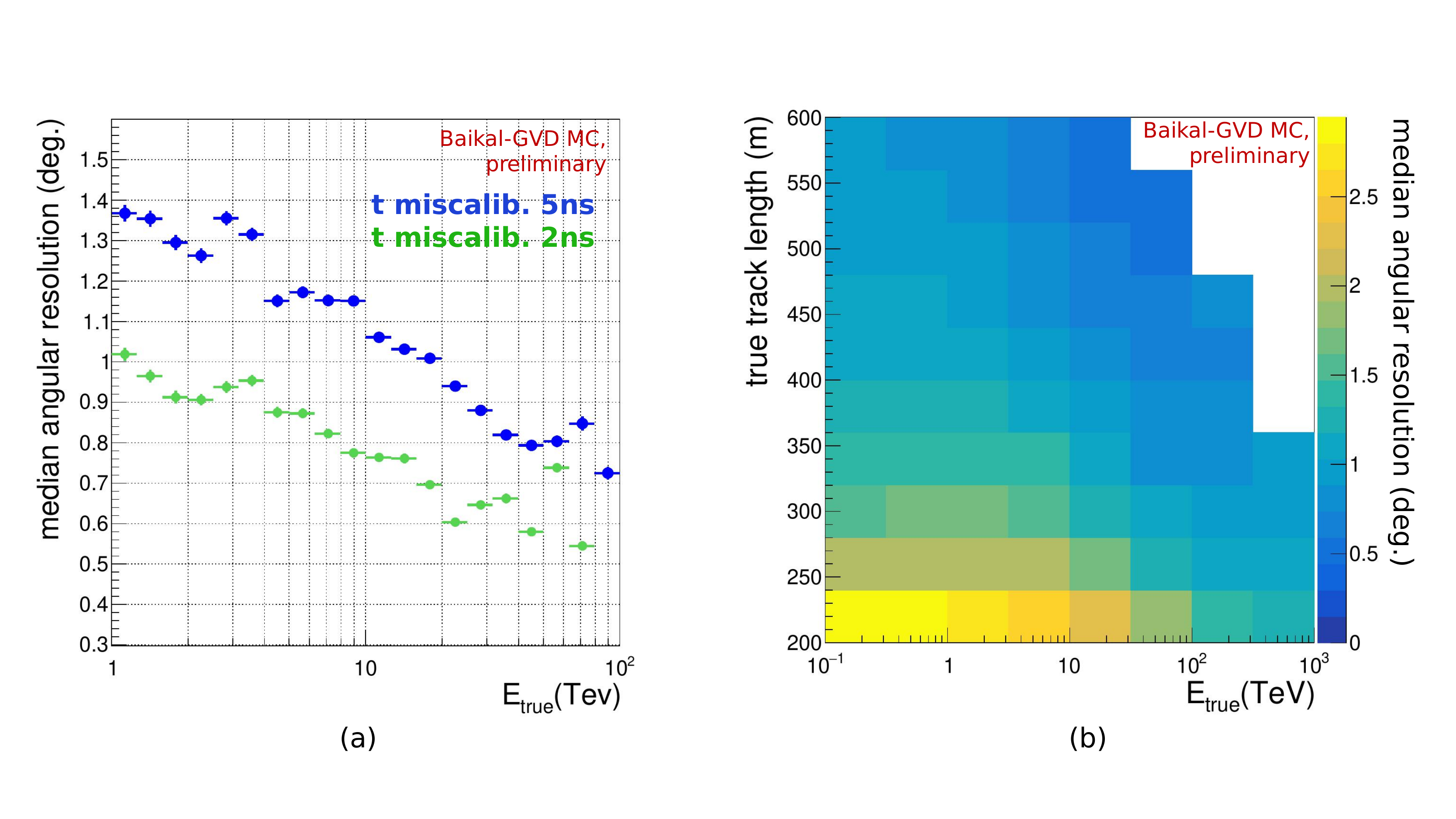}
  \vspace{-10pt}
  \caption{(a) Angular resolution attainable with the $\chi^{2}$-based track
reconstruction using a single Baikal-GVD cluster shown as a function of
neutrino energy for two different assumptions for the channel time
miscalbration: 2 ns (shown in green) and 5 ns (blue). (b) the same as
(a) shown as a function of neutrino energy and track length, assuming a
5 ns time miscalibration.} 
\label{fig:res_angular}
\end{figure}

An improved muon reconstruction employs the efficient hit finding algorithm developed for Baikal-GVD~\cite{scf}. This algorithm is based on the directional causality criteria developed in ANTARES experiment~\cite{bruijn}. Zenith and azimuthal angle space is scanned on the grid with fixed step size and for each direction a group of hits is selected. Large group of hits with the best quality criteria is selected as an input to track fitting algorithm. The scan grid is constructed around preliminary direction which is found using the pre-fit (1). The step size of 10$^{\circ}$ is used and $\pm60^{\circ}$ region around preliminary direction is scanned. The new hit finder algorithm provides sample of hits with $5\%$ noise contamination for low energy neutrino decreasing to $1-2\%$ for high energies \cite{scf}. At the same time the signal pulse selection efficiency is increased by the factor of 2-3 with respect to the fast reconstruction depending on muon energy and zenith angle. The track fitter is using $\chi^{2}(t)$ minimisation based on bare muon model. Thanks to directional causality criteria the new hit finder algorithm provides a sample of hits dominated by those falling well into the bare muon model disregarding late hits due to stochastic energy losses rate of which increases with muon energy. This property enables good $\chi^2(t)$ fit quality in wide energy range. The precision of track reconstruction depends on the time measurement precision. Channel-by-channel random gaussian time smearing modeling miscalibration of measurement channels is applied to MC. Miscalibration with $\sigma = 5$~ns results in resolution degradation by $\sim50\%$ with respect to miscalibration with $\sigma = 2$~ns (Fig.~\ref{fig:res_angular} (a)). Presently the $\sigma = 5$~ns miscalibration is applied to MC to attain better data-MC agreement in time residual distributions for PMT pulses selected for reconstruction. Median angular resolution depends on the track length and muon energy for shorter low-energy tracks the angular resolution can be as low as ~2.5$^{\circ}$ while for long tracks and $\sim100$~TeV energies the resolution imporoves to ~0.5-0.7$^{\circ}$ (Fig.~\ref{fig:res_angular} (b)).

\begin{figure}[ht]
\vspace{-10pt}
\centering
  \includegraphics[width=0.9\textwidth]{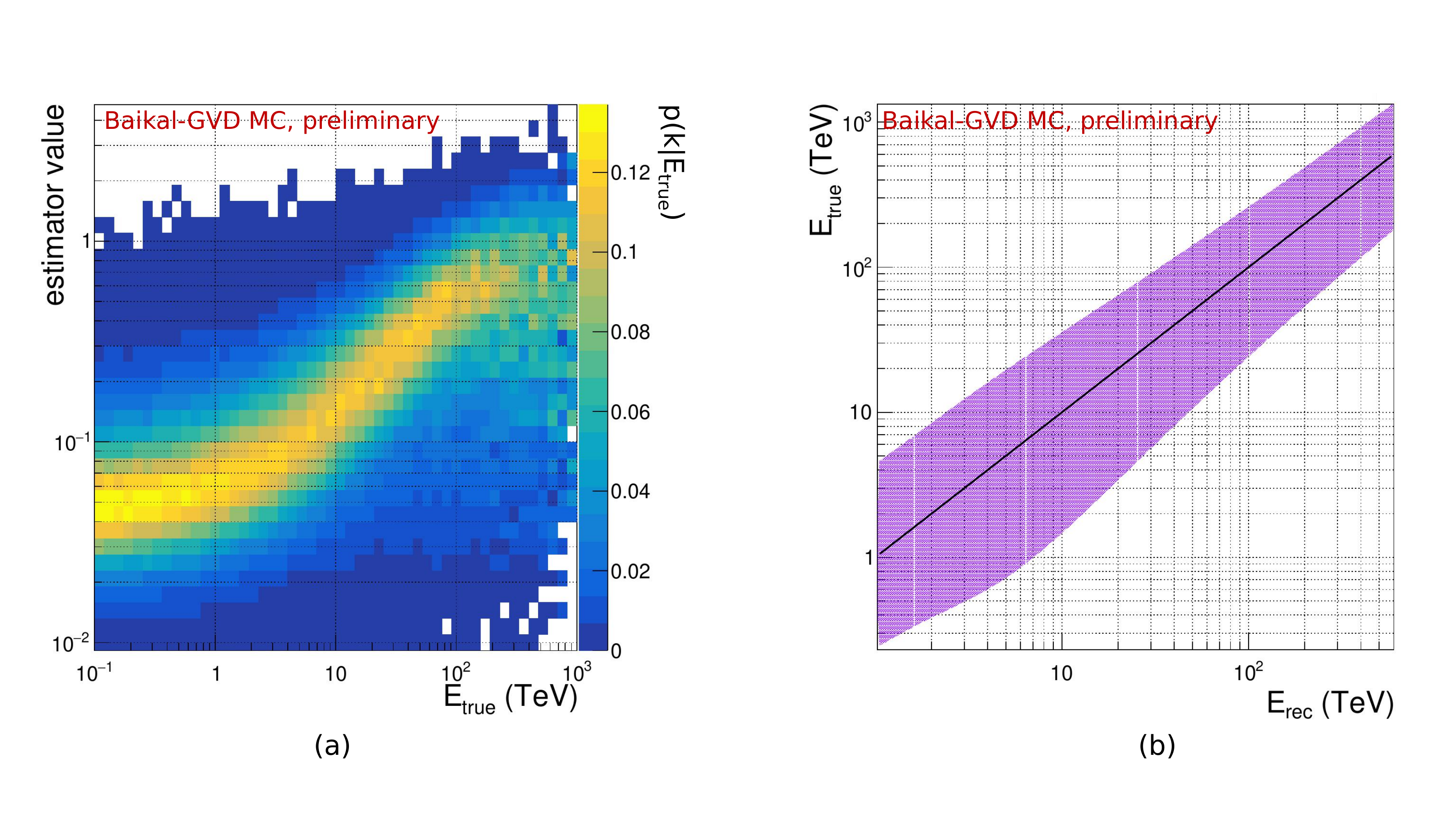}
  \vspace{-10pt}
  \caption{(a) Two-dimensional distribution showing the correlation of the energy
estimator value with the true muon energy, obtained using the
high-energy neutrino MC sample. (b) The median value (black line) and
the 68\% containment region (purple band) of the true muon
energy as a function of the reconstructed muon energy.}
\label{fig:res_energy}
\end{figure}

The energy of the muon is proportional to its average energy loss, $dE/dx$, starting from $\sim1$ TeV energy. Thus energy loss measurement can be used as muon energy estimator. An estimator for the muon energy, $k$, is constructed as follows:
\begin{equation}
k = {1 \over \epsilon L} \sum_{N_{hits}}q_i, \;\;\;\; \epsilon = \sum_{N_{ch}}{e^{-d/\lambda_{att}}\alpha(\Theta) \over d},
\end{equation} 
where $L$ is track intercept with the cylinder of cluster size with attenuation length, $\lambda_{att}=22$~m,  added to radius, top and bottom, $q_i$ is the charge deposited at $i$-th channel, $d$ - distance along the Cerenkov light cone from the track to OM, and $\alpha(\Theta)$ is angular sensitivity for given angle of incidence of Cerenkov photons onto the OM photocathode. The $\epsilon$ is a measure of detector sensitivity to the track. The clear correlation of $k$ and true muon energy, $E_{true}$ in demostrated in range $1<E_{true}<~150$~(TeV) for single-cluster analysis (Fig.~\ref{fig:res_energy} (a)). The mapping of $k$ to reconstructed muon energy, $E_{rec}(k)$, is obtained with the MC sample of neutrino with $E^{-2}$ spectrum by fitting median of $E_{true}$ distribution in bins of $k$ with 4th degree polynome. An uncertainty of $E_{rec}$ is obtained by fitting $E_{true}$ values defined by varying cumulative probability by $\pm0.341$. At Fig.\ref{fig:res_energy} (b) the $68\%$ confidence interval for $E_{true}$ is shown as a function of $E_{rec}$. The precision of energy reconstruction for 100 TeV muons amounts to factor $\sim$3.   
 
\section{Neutrino selection}
\begin{figure}[ht]
\vspace{-10pt}
\centering
  \includegraphics[width=0.9\textwidth]{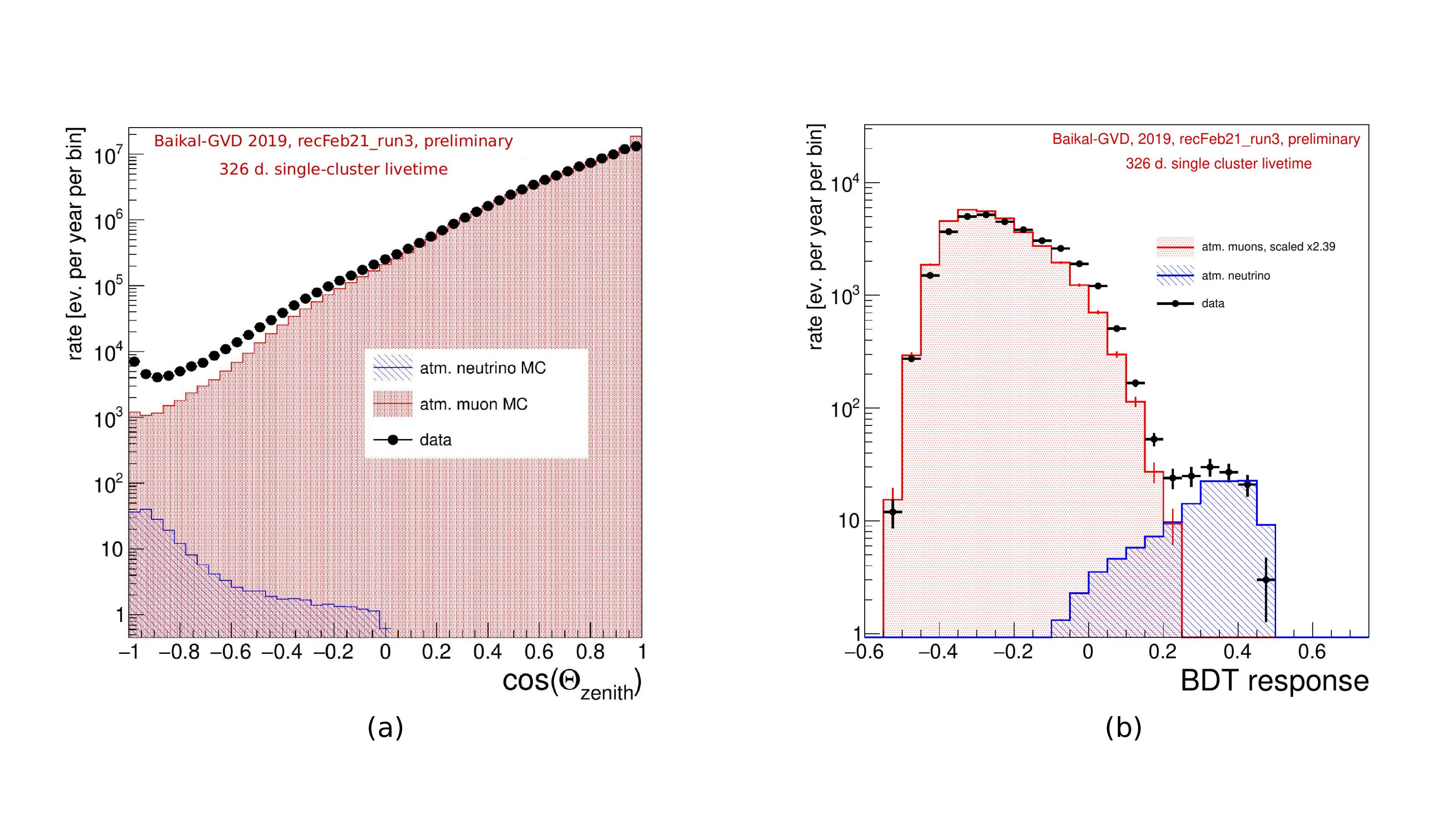}
  \vspace{-10pt}
  \caption{Reconstructed muon zenith angle distribution for data, atmospheric muon bundle MC and atmospheric neutrino MC (a). Events from different clusters are summed. In MC each cluster is summed with weight proportional to its livetime in data. Output of the BDT classifier for the data, atmospheric muon bundles and atmospheric neutrino (b). Muon bundle sample is scaled by 2.39 to match the normalisation of the data.}
\label{fig:muons}
\end{figure}

Previously we have developed a cut-based analysis which was applied to quiet period of 2019 and 44 neutrino candidates were selected \cite{atm_neutrino,nuatm_arxiv}. In this report we present complimentary neutrino selection method based on boosted decision tree (BDT) classifier which is applied to new reconstruction algorithm output on a sample of 326 days of single-cluster livetime for the same period. Root TMVA framework was used for BDT implementation and training \cite{tmva}. The BDT classifier with adaptive boost was selected. In this report we present low-energy neutrino selection while the high-energy neutrino selection algorithm is in preparation. 
Zenith angle distribution of reconstructed muon tracks for data, atmospheric muon bundle and atmspheric neutrino MC samples is shown at Fig.~\ref{fig:muons} (a). A good agreement of data and MC in downgoing region ($\cos(\Theta_{zenith})>0$) is observed while for upgoing region the rate of events in data exceeds MC prediction by up to factor $\sim$5. In single-cluster analysis the neutrino selection is developed for region $\Theta_{zenith}>120^{\circ}$. For this region the misreconstructed muon background needs to be suppressed by the factor of $10^3-10^4$. A set of loose quality cuts was applied both to signal and background samples before the BDT optimisation. A loose track fit quality cut was applied, $\chi^2/NDF < 30$, also a cut on the probability for the given collection of hits to fire $p_{hit}>10^{-8}$ was applied along with some other cuts. The signal efficiency for loose quality cuts is 88\% while misreconstructed muons are suppressed by $\sim$50\%. 
A set of 15 variables was used at the BDT input. These are the $N_{hits}$, $N_{strings}$, $\chi^2/NDF$, $\log_{10}(p_{hit})$, maximum distance between OM projection on the track: $Z_{dist}$, track isolation - fraction of empty OM in a 20 m radius around the track, maximum gap between OM projections along the track and other variables. The distribution of the BDT output response, $R_{BDT}$, for data, signal and background is shown at Fig.~\ref{fig:muons} (b). At this figure the background is scaled by 2.39 to match the total normalisation for data and partially compensate disagreement observed at Fig.~\ref{fig:muons} (a). The signal events form distinctive peak in the right part of the distribution. To select the neutrino event a cut of $R_{BDT}>0.25$ is applied. The efficiency of this cut for signal sample is $70\%$. No single MC muon backgound sample event out of 643 days of exposition survives this cut. 

\begin{figure}[ht]
\vspace{-10pt}
\centering
  \includegraphics[width=0.9\textwidth]{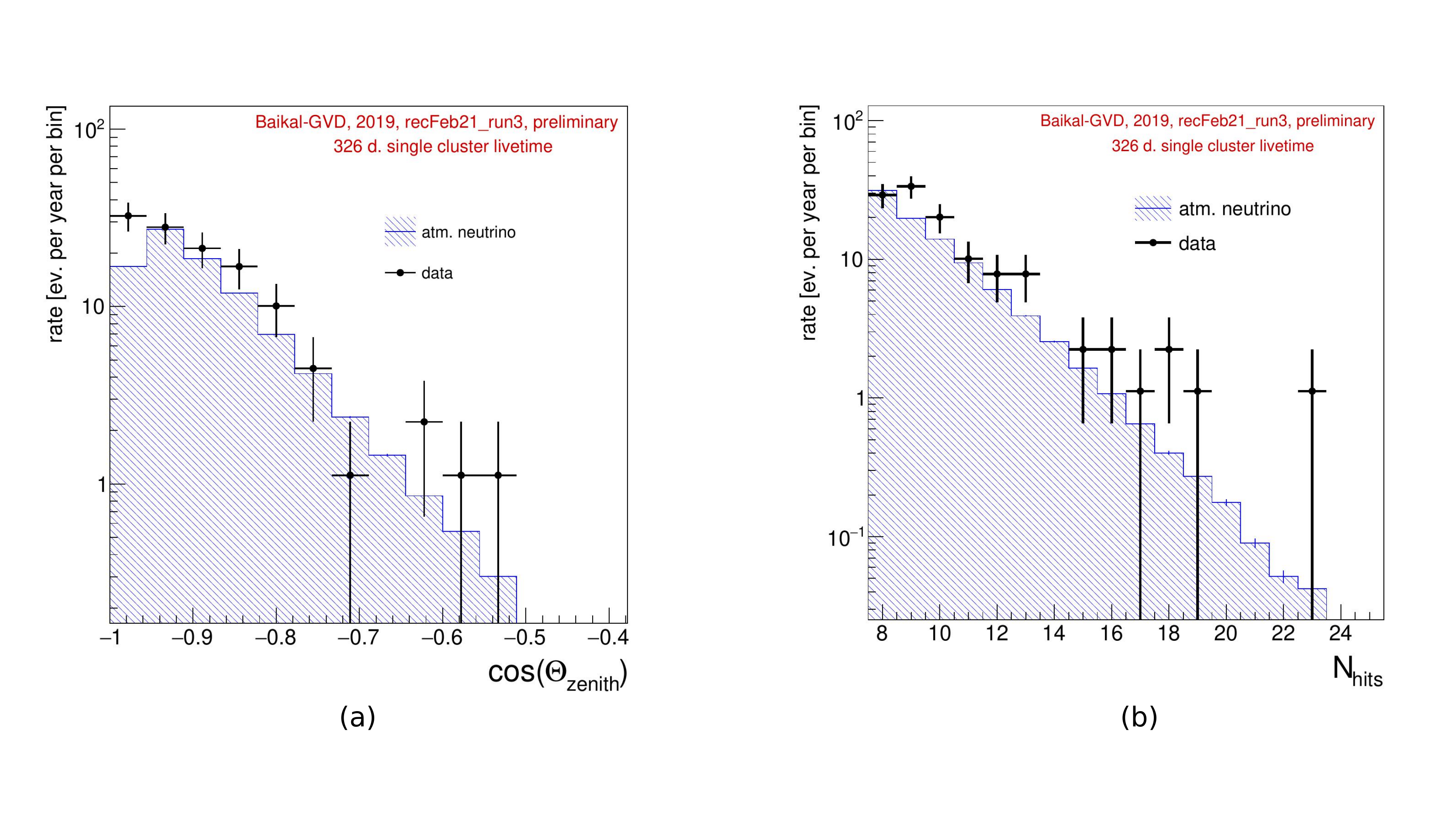}
   \vspace{-10pt}
   \caption{Distributions of reconstructed $\Theta_{zenith}$ (a) and $N_{hits}$ (b) for events selected with loose quality cuts and cut $R_{BDT}>0.25$.}  
\label{fig:neutrinos}
\end{figure}

After the $R_{BDT}>0.25$ cut 106 neutrino candidate events are selected in data while 81.2 events are expected from atmospheric neutrino MC. This suggests that there is $\sim 30\%$ disagreement between the data and MC which requires further study. At Fig.~\ref{fig:neutrinos} distributions of $\cos(\Theta_{zenith})$ and $N_{hits}$ for selected events are shown. The largest event observed in this sample is 23-hit event with almost vertical track ($\Theta_{zenith}=175.8^{\circ}$) while the charge deposition for this event is moderate. The median energy estimate for the most energetic event in this sample is $E_{rec}=28.3$ TeV while the 68\% confidence interval for the muon energy is $5.16 < E_\mu < 85.13$~(TeV). Despite the 30\% disagreement between the MC expectation and data in neutrino candidate rate presented analysis demonstrates that factor $~\sim$2 improvement of neutrino detection efficiency is possible with the new muon reconstruction algorithm in single-cluster analysis.    


\section{Conclusions}
Low-energy muon reconstruction algorithm is developed and deployed into the automatic data processing chain. First sample of low-energy neutrino candidates was selected with cut-based analysis. An extension of the muon reconstruction for large muon energies is in development. An improved reconstruction algorithm provides median angular resolution of $\sim0.5-0.7^{\circ}$ for 100 TeV muons. A factor $\sim$3 energy resolution is achieved for 100 TeV muons. Improvement of low-energy neutrino detection efficiency by the factor of $\sim$2 was demonstrated with BDT-based analysis. Sample of 106 neutrino candidates was selected with single-cluster analysis for April-June 2019 using the new reconstruction. An effort to extend muon data analysis to multi-cluster dataset and larger time span is ongoing.      

The work was partially supported by RFBR grant 20-02-00400. The CTU group acknowledges the support by European Regional Development Fund-Project No. CZ.02.1.01/0.0/0.0/16$\_$019/0000766. We also acknowledge the technical support of JINR staff for the computing facilities (JINR cloud).

\clearpage
\section*{Full Authors List: \Coll\ Collaboration}

%
\scriptsize
\noindent
{V.A.~Allakhverdyan}$^1$,
{A.D.~Avrorin}$^2$,
{A.V.~Avrorin}$^2$,
{V.M.~Aynutdinov}$^2$,
{R.~Bannasch}$^3$,
{Z.~Barda\v{c}ov\'{a}}$^4$,
{I.A.~Belolaptikov}$^1$,
{I.V.~Borina}$^1$,
{V.B.~Brudanin}$^{1\dagger}$,
{N.M.~Budnev}$^5$,
{V.Y.~Dik}$^1$,
{G.V.~Domogatsky}$^2$,
{A.A.~Doroshenko}$^2$,
{R.~Dvornick\'{y}}$^{1,4}$,
{A.N.~Dyachok}$^5$,
{Zh.-A.M.~Dzhilkibaev}$^2$,
{E.~Eckerov\'{a}}$^4$,
{T.V.~Elzhov}$^1$,
{L.~Fajt}$^6$,
{S.V.~Fialkovski}$^{7\dagger}$,
{A.R.~Gafarov}$^5$,
{K.V.~Golubkov}$^2$,
{N.S.~Gorshkov}$^1$,
{T.I.~Gress}$^5$,
{M.S.~Katulin}$^1$,
{K.G.~Kebkal}$^3$,
{O.G.~Kebkal}$^3$,
{E.V.~Khramov}$^1$,
{M.M.~Kolbin}$^1$,
{K.V.~Konischev}$^1$,
{K.A.~Kopa\'{n}ski}$^8$,
{A.V.~Korobchenko}$^1$,
{A.P.~Koshechkin}$^2$,
{V.A.~Kozhin}$^9$,
{M.V.~Kruglov}$^1$,
{M.K.~Kryukov}$^2$,
{V.F.~Kulepov}$^7$,
{Pa.~Malecki}$^8$,
{Y.M.~Malyshkin}$^1$,
{M.B.~Milenin}$^2$,
{R.R.~Mirgazov}$^5$,
{D.V.~Naumov}$^1$,
{V.~Nazari}$^1$,
{W.~Noga}$^8$,
{D.P.~Petukhov}$^2$,
{E.N.~Pliskovsky}$^1$,
{M.I.~Rozanov}$^{10}$,
{V.D.~Rushay}$^1$,
{E.V.~Ryabov}$^5$,
{G.B.~Safronov}$^2$,
{B.A.~Shaybonov}$^1$,
{M.D.~Shelepov}$^2$,
{F.~\v{S}imkovic}$^{1,4,6}$,
{A.E. Sirenko}$^1$,
{A.V.~Skurikhin}$^9$,
{A.G.~Solovjev}$^1$,
{M.N.~Sorokovikov}$^1$,
{I.~\v{S}tekl}$^6$,
{A.P.~Stromakov}$^2$,
{E.O.~Sushenok}$^1$,
{O.V.~Suvorova}$^2$,
{V.A.~Tabolenko}$^5$,
{B.A.~Tarashansky}$^5$,
{Y.V.~Yablokova}$^1$,
{S.A.~Yakovlev}$^3$
and
{D.N.~Zaborov}$^2$
\noindent

$^1$\textit{Joint Institute for Nuclear Research, Dubna, Russia}

$^2$\textit{Institute for Nuclear Research, Russian Academy of Sciences, Moscow, Russia}

$^3$\textit{EvoLogics GmbH, Berlin, Germany}

$^4$\textit{Comenius University, Bratislava, Slovakia}

$^5$\textit{Irkutsk State University, Irkutsk, Russia}

$^6$\textit{Czech Technical University in Prague, Prague, Czech Republic}

$^7$\textit{Nizhny Novgorod State Technical University, Nizhny Novgorod, Russia}

$^8$\textit{Institute of Nuclear Physics of Polish Academy of Sciences (IFJ~PAN), Krak\'{o}w, Poland}

$^9$\textit{Skobeltsyn Institute of Nuclear Physics, Moscow State University, Moscow, Russia}

$^{10}$\textit{St.~Petersburg State Marine Technical University, St.Petersburg, Russia}

\note[$\dagger$]{Deceased}

\end{document}